\documentclass[12pt,preprint]{aastex}

\def\Mdot{\hbox{$\dot {M}$}}

\def\Msun{\hbox{\it M$_\odot$}}
\def\Minit{\hbox{\it M$_{\rm initial}$}}
\def\Msunyr{\hbox{\it M$_\odot\,$yr$^{-1}$}}
\def\pc{\hbox{\it pc}}
\def\Myr{\hbox{\it Myr}}
\def\Gyr{\hbox{\it Gyr}}

\def\AK{\hbox{\it A$_{\rm K}$}}
\def\mnh{\hbox{\it m$_{\rm F160W}$}}
\def\mnk{\hbox{\it m$_{\rm F205W}$}}
\def\mK{\hbox{\it K}}
\def\mV{\hbox{\it V}}
\def\mH{\hbox{\it H}}
\def\mKp{\hbox{\it K'}}

\def\simgr{\mathrel{\hbox{\rlap{\hbox{\lower4pt\hbox{$\sim$}}}\hbox{$>$}}}}
\def\simls{\mathrel{\hbox{\rlap{\hbox{\lower4pt\hbox{$\sim$}}}\hbox{$<$}}}}

\makeatletter
\def\jnl@aj{AJ}
\ifx\revtex@jnl\jnl@aj\fi
\makeatother

\received{}
\revised{REVISION DATE}
\accepted{}
\journalid{VOL}{JOURNAL DATE}
\articleid{START PAGE}{END PAGE}
\paperid{MANUSCRIPT ID}
\cpright{TYPE}{YEAR}
\ccc{CODE}

\shorttitle{HST/NICMOS Observations}
\shortauthors{Figer et al.}

\begin{document}

\title{An Extended Star Formation History for the Galactic Center from 
Hubble Space Telescope/NICMOS Observations\footnote{Based on observations with the NASA/ESA Hubble Space Telescope, obtained
at the Space Telescope Science Institute, which is operated by the Association of Universities
for Research in Astronomy, Inc. under NASA contract No. NAS5-26555.}
}

\author{Donald F. Figer\altaffilmark{2,3}, R. Michael Rich\altaffilmark{4}, \\
Sungsoo S. Kim\altaffilmark{5}, Mark Morris\altaffilmark{4}, Eugene Serabyn\altaffilmark{6}}

\email{figer@stsci.edu}

\altaffiltext{2}{Space Telescope Science Institute, 3700 San Martin Drive, Baltimore, MD 21218; figer@stsci.edu }
\altaffiltext{3}{Department of Physics and Astronomy, Johns Hopkins University, Baltimore, MD  21218}
\altaffiltext{4}{Department of Physics and Astronomy, University of California, Los Angeles, 
	Division of Astronomy, Los Angeles, CA, 90095-1562 }
\altaffiltext{5}{Dept. of Astronomy \& Space Science, Kyung Hee University, Yongin-shi 449-701, Korea}
\altaffiltext{6}{Caltech, 320-47, Pasadena, CA 91125; eserabyn@huey.jpl.nasa.gov}

\begin{abstract}
We present Hubble Space Telescope (HST) 
Near-Infrared Camera and Multiobject Spectrometer (NICMOS) observations as evidence 
that continuous star formation has created much of the central stellar cusp of the Galaxy. 
The data are the deepest ever obtained for a Galactic Center (GC) population, 
being $>$50\% complete for \mnk$<19.3$, or initial stellar masses $\gtrsim$2~\Msun.
We use Geneva and Padova stellar evolution models to produce synthetic luminosity
functions for burst and continuous star formation scenarios, finding that the
observations are fit best by continuous star formation at a rate that is
consistent with the recent star formation activity that produced the three
massive young clusters in the central 50~\pc. Further, it is not possible to
fit the observations with ancient burst models, such as would be appropriate
for an old population like that in Baade's Window or NGC6528.  
\end{abstract}


\section{Introduction}
The Milky Way Galaxy is composed of several distinct stellar components, including the disk, the
bulge, and the halo. In addition to these well-known populations, new
evidence suggests that there is yet another stellar population in the central
part of the bulge, comprised of both young and old stars. The old component has often
been associated with the inward extension of the ancient bulge, and the young stars
are generally clustered in the central parsec and in two massive clusters about 30~\pc\ to the
north of the center. Stars in the central region are currently forming in several
molecular clouds, i.e.\ Sgr~B2.
On scales covering a few hundred parsecs, \citet{ser96} identify a peak in the infrared
surface brightness distribution, scaling as 1/r in COBE data, associating this feature with light from
a massive cluster of stars having a 1/r$^2$ volume density distribution, 
the ``r$^{-2}$'' cusp \citep{hir84}. \citet{ser96} suggest that the 
cusp is composed of stars created continuously over the lifetime of 
the Galaxy. \citet{lau02} find a central disk in the
COBE imagery of the Galactic center (GC), and identify this component with the central cusp.

\citet{cat90} resolve this cusp (over degree scales) into individual bright stars, and argue that the
overall stellar population was likely to be intermediate aged, with the youngest
population more concentrated toward the center. Others find similar evidence in
support of an intermediate-age stellar population in the Galactic Center (mostly in the 
central few arcminutes), i.e.\ 
\citet{rie93}, \citet{hal92}, \citet{hal89}, \citet{rie87}, 
\citet{leb87}, \citet{blu95}, \citet{nar96}, and \citet{sjo99}. \citet{fro99} present some of the
strongest evidence associating the r$^{-2}$ cusp with an intermediate population,
finding that the density of young stars near the GC declines much more
rapidly with galactocentric radius than does the density of the ancient bulge population. 

The evidence for very recent ($<$10~\Myr) star formation in the GC abounds.
The Lyman continuum flux emitted in the central few degrees of the Galaxy is
$\sim$10$^{52}$ photons/s \citep{cox89}, half coming from stars in the
three massive young clusters in the central 50~\pc\ \citep{fig99a,fig02}. 
This represents about 10\% of the total Lyman continuum flux for the whole Galaxy, and 
the number of massive stars (\Minit$>$20~\Msun\ in the context of this paper) 
in the clusters is also about 10\% of the
number in the whole Galaxy. However, the star formation rate in the GC is $\sim$1\% 
of the total star formation rate in the Galaxy, as judged by simply dividing the mass in known, 
recently formed, stars by the duration of the star formation episodes that formed those
stars, i.e.\  5(10$^4$)~\Msun/5~\Myr$\sim$0.01~\Msunyr,
giving a star formation rate density of 10$^{-7}$~\Msunyr~\pc$^{-3}$. This rate is roughly a factor of 
250 higher than the mean rate in the Galaxy, and about the same factor lower than
the rate in starburst galaxies. Such a low star formation rate compared to 
Lyman continuum photon production necessarily
follows from the relatively flat initial mass function (IMF) slope estimated by
\citet{fig99b} and used in estimating the mass of stars formed in the young clusters. 

The young stellar clusters in the GC are extraordinary. 
The Central cluster is located within the central parsec and contains over 30 massive stars \citep{gen96}. 
Only 30~\pc\ distant, by projection, from
the center are the Arches \citep{cot94,fig95a,nag95,cot95,cot96,ser98,fig99b,fig02} and Quintuplet clusters 
\citep{oku90,nag90,gla90,mon92,mon94,geb94,fig95b,fig99a,fig99b}. The 
Arches cluster contains at least 150 O-stars within a diameter of 0.6~\pc, 
making this cluster the densest in the Galaxy \citep{fig99b}. A list of the 
30 most massive stars in the Quintuplet cluster is given in \citet{fig99a}. All three clusters
are quite similar in most respects, except for age; the Central and Quintuplet clusters are $\approx$
3$-$5 Myr old \citep{fig99a,kra95,naj97}, while the Arches cluster is substantially 
younger, $\tau_{\rm age}~=~2.5\pm$0.5~\Myr\ \citep{fig02}. 
With such a collection of young, massive clusters, is it possible that we are witnessing
an extraordinary burst of star formation in the GC, or has the center hosted similar
bursts of star formation throughout its long history? Further, what is the fate of these
massive clusters?

\citet{kim99}, \citet{kim00}, and \citet{zwa01} argue that
the young clusters in the GC will be tidally disrupted in $\simls$~10$-$50~\Myr. This
might explain the absence of older clusters of similar masses in this region. 
\citet{ger01} argues that such clusters should spiral into the central parsec
on relatively short timescales, i.e.\ that the HeI stars in the central cluster might contain
stellar products of a dense cluster formed well outside the central parsec. In this
scenario, the stars are drawn inward to the central parsec by dynamical friction between
the cluster and the tidal field of the GC. \citet{kim03a} and \citet{kim03b} explore this possibility
using N-body simulations, finding that dynamical friction is unlikely to have produced
the population presently seen in the central parsec.

Clearly, the GC has formed a plethora of stars in the past
5~\Myr, but it is less apparent when the bulk of stars in the central
50~\pc\ formed. If we assume that the star formation rate in the past was similar
to the present rate, then the total mass of stars formed over the past 10~\Gyr\ is
$\approx$10$^8$~\Msun\ within a radius of 30~\pc\ of the GC, 
or an order of magnitude greater than this amount over the
whole Central Molecular Zone, as first suggested by Serabyn \& Morris (1996). 

This paper reports new, deep imaging of several fields in the central 100~\pc\ of the Galaxy.
Our imaging solidly reaches the helium-burning clump giant stars and 
just reaches the old main sequence turn-off point. We model the 
luminosity functions to explore the star formation history of the Galactic center.
We argue that the best fit to the total number of stars and to the luminosity functions
requires an approximately continuous star formation history over the last $\sim$10~\Gyr.

\section{Observations}

The Galactic Center observational data are taken from \citet{fig95a}, 
Gemini/AO science verification observations\footnote{Based on observations obtained at 
the Gemini Observatory, which is operated by the Association of Universities for 
Research in Astronomy, Inc., under a cooperative agreement with the NSF on behalf of 
the Gemini partnership: the National Science Foundation (United States), the Particle 
Physics and Astronomy Research Council (United Kingdom), the National Research 
Council (Canada), CONICYT (Chile), the Australian Research Council (Australia), 
CNPq (Brazil) and CONICET (Argentina).}, and NICMOS observations obtained
as a part of HST program GO-7364. The field locations are shown in Figure~\ref{fig-fields} and
listed in Table~1. 

The Lick data cover 40 fields, in a mosaic, obtained at the Shane 3-m telescope at Lick Observatory
in the \mH\ ($\lambda_{\rm center}$=1.65~\micron) 
and \mKp\ ($\lambda_{\rm center}$=2.12~\micron) filters. The individual images cover
3$\arcmin \times 3\arcmin$ of area, while the mosaic spans 24$\arcmin$ $\times$ 14\arcmin\ of
area oriented with the long axis north-south and centered 6 $\arcmin$ to the north of the GC. The exposure
times were 35~seconds for each image.

The Gemini/AO data were obtained as part of the Gemini North
commissioning observing run using the Hokupa'a+Quirc instrument in \mH- and \mKp-bands. The field
size is 20$\arcsec \times 20\arcsec$.
These data cover 4 fields (\#1, 2, 5, and 6), arranged in a mosaic centered about 15\arcsec\ to the north and east 
of the GC. Both the ``short'' ($\tau_{\rm exposure}$=1 second)
and ``long'' ($\tau_{\rm exposure}$=30 seconds) exposure images were used. 

The NICMOS data include 12 fields obtained with the NIC2 camera (19\farcs2 on a side) in
the F110W ($\lambda_{\rm center}$=1.10~\micron), F160W ($\lambda_{\rm center}$=1.60~\micron), 
and F205W filters ($\lambda_{\rm center}$=2.05~\micron). 
We selected these fields based on a number of criteria: avoidance of
crowded regions (thus excluding the centralmost region), avoidance of
young star clusters, avoidance of regions of high extinction, and
inclusion of fields ranging over a variety of Galactic latitudes and
longitudes. Because extant near-IR maps of the Galactic nucleus, i.e.\ from 
2MASS, show extensive regions of deep extinction just above and below
the actual GC, our observations consisted mainly of a
vertical strip of positions at a Galactic longitude offset, $\Delta l$, of 12' from our Galaxy's central radio point source Sgr A$^*$ (our
``b strip''), as well as one field located along the true galactic
plane, at half the longitude offset of our vertical strip (our l/2
position). Observation of these undistinguished ``background'' fields
in the nuclear cluster was obtained in concert with our observations
of the young ``Arches'' and ``Quintuplet'' clusters. Those cluster observations also included
four ``off'' positions surrounding each cluster. The exposure
times were 255~seconds for each image.

We use aperture photometry with d$_{\rm aperture}$=4$\farcs$2, or 6 pixels, for
the Lick data set, with zero-point calibration set by photometry of several bright
stars in the mosaic area from the literature. We used PSF-fitting photometry for both the 
HST/NICMOS and Gemini/AO data sets. Zero-point calibration for the HST/NICMOS 
data was taken from \citet{mac97}, with specific details relevant to our data set, including completeness corrections,
discussed in \citet{fig99b}.
The Gemini/AO zero-point calibration was set by photometry of several stars in \citet{blu96}.
The Gemini and Lick \mKp\ data were converted into \mK\ data using the observed \mH$-$\mKp\
values and the relation of \citet{wai92}.

\section{Analysis and Results}

\subsection{Color-magnitude Diagrams}
Color-magnitude diagrams (CMDs) for the three data sets are shown in Figure~\ref{fig-panels_cmd}.
We can see that the NICMOS data continue
down to about \mnk=21 (\Minit$\sim$1.2~\Msun), roughly the magnitude for which the S/N$\sim$5. This is
about 4 magnitudes fainter than the Gemini/AO data, and about 1 magnitude fainter than the
data in \citet{gen03}. The data are complete at the 50\% level at \mnk=19.3 (\Minit$\sim$2.2~\Msun), averaged
over all fields. The Gemini/AO data go just faint enough to distinguish the ``red
clump'' near \mK$\sim$15.5. The Lick data are already incomplete at \mK$\sim$12, but they cover a much larger
area than the other data sets, giving us more complete statistics at bright magnitudes. 

CMDs for the individual NICMOS fields are shown in Figure~\ref{fig-individualcmds}. 
The trend in extinction for these fields is as expected in that the fields closest to
the Galactic Center suffer the greatest extinction, e.g., the zd field. Apparently,
stars in the za, qc, and ac fields are comingled with ambient molecular material that serves to
spread their locations in the diagram; the same effect is observed in the Lick
data for this location. The red clump population is prominent in all of the fields, although the
feature in the zc panel is weak. Differential reddening is prominent in each field and
broadens the clump distribution in color along the direction of reddening (to the lower right). 
The rise in the luminosity function at the main sequence turnoff point (e.g.\ at \mnk$\simls$19 and
\mnh$-$\mnk$\sim$1.6 in the za field) is seen clearly in all
fields, except for zd (closest to the Galactic Center) where confusion sets in at brighter
magnitudes. 

 The presence of the clump and the main
sequence turnoff in the luminosity function makes a prominent gap in the stellar CMD, an indication that 
much of the population must be older than a few \Gyr.  The relatively wide
$(\sim 4~mag)$ gap between the clump and main sequence
turnoff seems consistent with a population older than 10 \Gyr. 
As we proceeded with this interpretation, however, we tried to fit
old globular cluster luminosity functions to our Galactic Center data and
noticed that the red clump was too bright to be consistent
with a purely old stellar population.  After considerable rechecking and verification of 
our photometry, this puzzle led
us to proceed with the analysis we report in this paper.

\subsection{Luminosity Functions}
In order to compare the observations to model predictions, we 
construct dereddened luminosity functions (LFs) by subtracting  
individual reddening values for each star based upon its color in
H$-$K, or \mnh$-$\mnk, and the assumption that each star 
has intrinsic colors of a red giant. We estimate the intrinsic colors of a typical red giant by
convolving the spectral energy distribution of a K4III star from the \citet{pic98}
library with the filter profiles, finding (H$-$K)$_{0}$=0.11 and 
(\mnh$-$\mnk)$_{0}$=0.24. 
We deredden the photometry using the redenning law of \citet{rie89}, i.e.\ 
A$_{\lambda}\propto\lambda^{-1.53}$. LFs for the individual NICMOS fields are 
shown in Figure~\ref{fig-individuallfs}.

Figure~\ref{fig-total_lf0.0.za.zb.zc.zd.qca.qcb.qcc.qcd.aca.acb.acc.the.gem.black} 
shows the combined dereddened luminosity functions for the NICMOS, Lick, and Gemini data. 
The ``NICMOS'' curve represents the sum
of all the NICMOS fields, and it is scaled to account for the much larger area
covered by the Lick survey. We find that (\mnk$-$\mK)$_0$$<$0.05 for
red clump stars, as measured by convolving the filter profiles with a spectral energy
distribution of such stars; we thus make no explicit correction 
when comparing photometry in the two filters. 
The Lick data are confusion limited for \mK$_{\rm 0}>$9, where the counts begin to roll over. 
The NICMOS luminosity function appears to be a straight line, except for a bump at \mK$_{\rm 0}$=12,
which represents the red clump.
The two luminosity functions appear to join for brighter magnitudes. 
The Gemini counts have been arbitrarily divided 
by 9 in order to scale them to match the NICMOS counts; the large difference in
observed surface number density is expected, given that the Gemini fields are located
much nearer to the GC than the NICMOS fields.
The Gemini, Lick, and NICMOS data match very well, except for the weakness of the red clump in
the Gemini data, a feature that is likely due to the incompleteness of the Gemini data.

\subsection{Star Formation Models}
We now infer the star formation history responsible for the observed populations by
comparing their observed luminosity functions to models built with both the Geneva 
and Padova isochrones. The Geneva models are described
in \citet{sch92,sch93a,sch93b,cha93,mey94}. They cover a grid of metallicities (Z$_\odot$/20, Z$_\odot$/5,
Z$_\odot$/2.5, Z$_\odot$, and 2Z$_\odot$), initial stellar masses (0.8~\Msun\ to 120~\Msun), and two
mass-loss rate laws for stars with \Minit$>$12~\Msun\ (the ``canonical'' and ``enhanced'' mass-loss rate laws).
We interpolate the evolutionary tracks and produce isochrones for a variety of masses and ages,
extrapolating below 0.8~\Msun. The Padova models were kindly provided by Leo Girardi, and based 
on work described in \citet{gir02}. They cover the same metallicities as
the Geneva models, and initial stellar mass between 0.15~\Msun\ and 80~\Msun. We do not use the Padova
isochrones to model starbursts that are young enough to still include stars with \Minit$>$80~\Msun\ (several
million years). 

The model luminosity functions for a given star formation history are produced by 
summing individual luminosity functions for separate
star formation events, under the constraint that the total mass in stars formed is
equal to 2(10$^{8}$)~\Msun, the mass inferred from velocity measurements in \citet{mcg89}
within a projected radius of 30~\pc\ of the Galactic Center. The individual 
events start at an age of 10$^7$ years and are separated by
10$^7$ years up to 10$^9$ years, at which point they are spaced by 10$^9$ years, up to
10$^{10.1}$ years.
The number of stars formed in each event determines the normalization of the
integral over a mass spectrum of stars given by a power-law with
index of $-$0.9, i.e.\ dN/dm=m$^{\Gamma}$=m$^{-0.9}$, as opposed to the ``Salpeter'' value of $-$1.35 \citep{sal55}; this
value is motivated by the measurements for the Arches cluster \citep{fig99b,yan02,sto03}. The upper mass limit
is 120~\Msun, and the lower mass limit is 0.1~\Msun. These masses are transformed to absolute magnitudes in
the \mV\ band through the models. These are converted to absolute magnitudes in the band of
interest, i.e.\ \mK, through a lookup table that relates color index to temperature (which comes
from the models). The apparent magnitude in the band of interest is then simply the
absolute magnitude plus the distance modulus (14.52, i.e.\ d=8~{\it kpc}). We then sum the histogram to
produce the luminosity function, and sum the individual luminosity functions to produce
the final luminosity function for a given star formation history. The normalization to apparent
surface density is fixed by dividing the model luminosity functions by an ``appropriate''
area, $\pi$(30~\pc)$^2$ throughout this paper. Note that this normalization produces a
single ``average'' density for the whole area, i.e.\ no attempt was made to scale the
densities with Galactocentric radius. 

Figure~\ref{fig-lfpanels0} shows a subset of starburst luminosity functions that we use in 
constructing the summed luminosity functions for various model star formation histories. 
In this figure, we illustrate a range of single starburst models, some of which are
ultimately used in constructing our more complicated star formation models histories.  
We consider a range of ages, assuming the Geneva models with Z=Z$_{\odot}$ and canonical mass-loss rates.
The Lick data are overplotted for comparison at the bright end (\mK$<$9), where the data
are reasonably complete. 
The luminosity functions for young starbursts have a gap between the red supergiant stars and
fainter stars that are still on the main sequence, a feature not seen in the Lick data.
For older starbursts, the supergiants are gone
and a horizontal branch/red clump forms (at $\sim$500~\Myr). Note that the red clump 
magnitude shifts to fainter, then brighter, over the time interval from 500~\Myr\ to 2~\Gyr.
This effect is most clearly demonstrated in the plot adapted from \citet{gir02} in Figure~\ref{fig-redclump}.
This figure agrees well with observed data of red clump stars for the ancient population
of Baade's Window, for which m$_{\rm RC}$=13.12 \citep{alv00}.

\subsection{Star Formation History}
The observations cannot be fit by any single starburst population. However, there
is promise in reproducing the observed features by constructing a model for episodic,
or continuous, star formation histories. 
We construct a model (Figure~\ref{fig-total_lf1.0c.za.zb.zc.zd.qca.qcb.qcc.qcd.aca.acb.acc.bw.black.cc}) 
by assuming that the GC region continuously produced stars that evolved according to the Geneva models \citep{mey94}. 
We assume that there was a burst every
10~\Myr\ from 10~\Gyr\ ago to the present, with an average rate of 0.07~\Msunyr. We choose
this rate so that the model curve fits the observed curve, but note that it is approximately
seven times the rate corresponding to the star formation that produced the three young clusters. 
The \citet{tie95} data of Baade's window stars were
arbitrarily scaled such that the counts roughly match the counts in the scaled
NICMOS data. The bright stars from the Lick data are well matched by our model.
Indeed, the bright end of the luminosity function can only be fit by stars younger than $\approx$100~\Myr.
Our model also matches the red clump at \mK$_0\sim$12 very well. Most importantly, the
model fits the absolute number of stars formed over the region modelled. 

Figures~\ref{fig-sfpanels1.0c.za.zb.zc.zd.qca.qcb.qcc.qcd.aca.acb.acc.cc} and
\ref{fig-sfpanels1.0x.za.zb.zc.zd.qca.qcb.qcc.qcd.aca.acb.acc.cc}
show alternate star formation scenarios, with the model counts modified by the observed
completeness fractions from \citet{fig99b}. The star formation histories are of three
families: ancient bursts, continuous star formation, and bursts plus continuous star formation. 
In all cases, the histories produce 2(10$^8)$~\Msun\ in stars within a radius of 30~\pc.

There are five constraints provided by the
models: 1) the counts in the bright end (\mK$_0<$8), 2) the slope at intermediate magnitudes
(10$<$\mK$_0<$15), 3) the
brightness of the red clump, 4) the counts at the faint end (\mK$_0>$15), and 5) the absolute counts 
per unit area. The counts in the bright
end are controlled by the extent of recent star formation. The slope is controlled by the
presence of a red giant branch; note that any star formation history that includes some
ancient stars will produce a red giant branch, and thus an intermediate magnitude slope that is a
relatively constant function versus specifics within that history. The brightness of the
red clump is related to
the extent of star formation activity at intermediate age. The
counts at the faint end are controlled by ancient star formation. The total number of counts
in each bin is controlled by the strength and overall age of the star formation. 
So, we find that we
can constrain the relative amounts of recent, intermediate, and ancient star 
formation activity through the use of these luminosity functions, in addition to the
absolute productivity of the star formation. 

Figures~\ref{fig-sfpanels1.0c.za.zb.zc.zd.qca.qcb.qcc.qcd.aca.acb.acc.cc} and 
\ref{fig-sfpanels1.0x.za.zb.zc.zd.qca.qcb.qcc.qcd.aca.acb.acc.cc} 
show that all five constraints are best fit by the continuous star formation
model. Indeed, the ancient bursts do not reproduce a bright end at all. The observed brightness of
the red clump is too bright for intermediate age bursts, whereas the continuous star formation
scenario fits this constraint well. The counts at the faint end are overpredicted in the
ancient burst models, but reasonably well fit by the continuous star formation model. Note that
the data are much more than 50\% incomplete for the faintest few bins. Most importantly, the absolute
numbers of stars at intermediate magnitudes cannot be reproduced by the ancient burst models. 
Indeed, the ancient burst models fail badly at
reproducing the number of stars seen, by two orders of magnitude in the brightest bins, 
even though the bursts assume a burst mass of 2(10$^8$)~\Msun.

The qualitative analysis above begs the question of uniqueness. In order to determine
the sensitivity of our technique, we attempt to model the old populations in Baade's
Window and in the Galactic globular cluster, NGC6528.
Because we do not have reliable
total mass constraints, unlike the situation for the GC, we scale the observed counts arbitrarily to 
achieve the best fit. Figures~\ref{fig-sfpanels1.0c.bw.auto} and \ref{fig-sfpanels1.0x.bw.auto} 
show the results for Baade's Window data. For Baade's Window, the ancient burst model provides the best
fit, reproducing the faintness of the red clump and the counts at the faint end. 
Note, however, that the number of
stars having 8$<$\mK$<$10 is not reproduced by the ancient burst models, because the models fail
to faithfully model the upper tip of the AGB luminosity function. 
The AGB is the most poorly understood evolutionary phase, and therefore
will be most poorly fit by any theoretical model.  Further, the AGB
luminosity has relatively low sensitivity to age, at Solar metallicity.
Finally, the lifetimes are short, so the number counts of stars tend
to be low.  For these reasons we do not place most of the weight
on the bright end of the AGB. The next best fit is provided
by the continuous star formation model, although the counts at the faint end and the
brightness of the red clump are not well reproduced.

Figures~\ref{fig-sfpanels2.0c.6528.auto} and \ref{fig-sfpanels2.0x.6528.auto} 
show the results for NGC6528. We use models for metallicity twice that of the
Solar value, in accordance with the measured abundances in the NGC6528 cluster stars \citep{car01}. 
Again, the figures show that the ancient burst model provides the best fit. 
The observed red clump is more pronounced than seen in the Geneva model, and 
it is very well fit by the Padova model, as is a second bump one magnitude fainter (\mK=14.0).
As in the case of Baade's Window, the next best fit is given by the continuous
star formation scenario. Just as before, this model for this scenario fails
to match the observed faint end and the brigthness of the red clump.

Next, we investigate the star formation history as a function of field location.
Figures~\ref{fig-sfpanels1.0c.za.cc}, \ref{fig-sfpanels1.0c.zb.cc}, \ref{fig-sfpanels1.0c.zc.cc}, 
and \ref{fig-sfpanels1.0c.zd.cc} show the completeness-corrected models overplotted on the
individual observations in the z-fields for the Geneva models, and 
Figures~\ref{fig-sfpanels1.0x.za.cc}, \ref{fig-sfpanels1.0x.zb.cc}, \ref{fig-sfpanels1.0x.zc.cc}, 
and \ref{fig-sfpanels1.0x.zd.cc} show the same for the Padova models. The luminosity
functions show that brighter stars are preferentially located closer to the Galactic
center and are absent for the zc field (\mnk$<$9). Note that this is not a sampling effect,
given that the fields are the same size. The brightness of the red clump is well-fit in
all cases, although the observed clump for the zd field extends over several bins. The overall
number of stars is also noticably elevated for the zd field. We suggest that all of
these effects are owed to more recent star formation closer to the Galactic center.
With greater areal sampling, we hope to constrain the star
formation history as a function of position in the Galactic center beyond the suggestive
variations we already see in the z fields. 

\subsection{Uniqueness}
We also ran models for a lower-mass cutoff (m$_{\rm lower}$) of 1~\Msun, instead 
of 0.1~\Msun. That primarily
resulted in a vertical shift upward of the luminosity functions for \mK$_0<$22. While this
elevated lower mass cutoff is consistent with the observations, we note that the absolute
vertical scale for the models is more uncertain that the difference seen between the two
cases of lower mass cutoffs. The results are also generally 
robust against variations in m$_{\rm upper}$, \Mdot, {\it Z}, and $\Gamma$, within a factor
of two in each parameter. 

\section{Discussion}

Although the presence of the red clump population is consistent with
a population older than 1~\Gyr, our modeling of the observed luminosity
function points strongly in the direction of a continuous star formation
history.  The number counts and shape of the luminosity function are
inconsistent with a population dominated by an ancient burst, or by
a small number of bursts older than 1~\Gyr.
We favor a continuous star formation history with a rate of $\sim$0.02~\Msunyr, or twice
the rate inferred by the presence of the bright young clusters in the region. This rate
appears to produce too few stars to match the observations, but it is bounded by the present
enclosed mass of 2(10$^8$)~\Msun\ within 30~\pc\ of the GC. A more refined estimate of the
rate will depend on careful normalization of the modelled surface number density. The young
clusters presently observed were formed at an average star formation rate of 0.01~\Msunyr, over the past 5~\Myr,
in good agreement with the conclusions of this paper. 

\subsection{Comparison to Other Work}
There is an abundant body of work noting the very young stellar population ($\tau_{\rm age}<10$~\Myr)
in the central few arcminutes, and additional studies noting intermediate-age populations
on size scales observed in this paper. In particular, \citet{nar96}, 
\citet{hal92}, and \citet{blu95} note an overabundance
of bright stars in this region, compared to Baade's Window. \citet{fro99} also note a dramatic increase in the
number of bright stars compared to Baade's Window. Our observations are consistent with these
studies in showing an increase in the number of bright stars from the furthest field (za)
to the closest field (zd) to the GC. Our results are consistent with previous
work finding evidence for an intermediate-age stellar population in the GC.
\citet{sjo99} identified a population of OH/IR stars having high wind velocities, 
suggesting a starburst $\sim 1$\Gyr\
ago. Our analysis rules out the possibility that the bulk of stars in the region formed
in such a burst, although the analysis could accommodate a modest sized burst at t=1~\Gyr,
if accompanied by continuous star formation at other times. 

\citet{gen03} published a K-band luminosity function for stars in the central
parsec, obtained with VLT/AO. Their Figure~9 shows the observations and a fit by a
an ancient single starburst model, noting the good agreement in the brightnesses of
the red clump and the overall slope, after accounting for young stars at the bright
end in the observations. It is somewhat difficult to compare our results with those
in \citet{gen03} for several reasons. First, the \citet{gen03} data are displayed in
one magnitude wide bins, potentially smoothing the effect of a bright red clump. Second,
the data are presented in the reddened reference frame with a single extinction value
being applied to the model. Our observations are first dereddened individually for
each star and then compared to the models. This subtle difference can affect the
detailed shape of the red clump in the case that differential extinction is important.
Finally, a detailed comparison would require that a single extinction law and
common wavebands be used in both cases. Indeed, we infer a larger extinction for
the central parsec than for our NICMOS fields, as shown in Figure~\ref{fig-extinction},
yet the extinction value in \citet{gen03} (\AK=3.2) is similar to the average
extinction values for our NICMOS fields. 

We do note, however, that our data produce a broad hump in the luminosity function between
\mK~10 and 13 in the reddened frame (Figure~\ref{fig-total_lf1.0c.za.zb.zc.zd.qca.qcb.qcc.qcd.aca.acb.acc.bw.black.cc}),
similar to that seen in Figure~9 of \citet{gen03}. The fact that we see the feature
in our data and in the \citet{gen03} data leads us to believe that the feature is
common to the central bulge, a result of ongoing star formation, and not solely 
a product of the very recent star formation in the central parsec. 

\subsection{Mass Budget} 
The molecular clouds in the GC provide the material that feeds the star formation at a 
rate of at a few hundredths of a solar mass per year \citep{mor01,fig02modes}.
The star formation occurs within the Central Molecular Zone (CMZ), a disk-like region 
of enhanced molecular density within a radius of $\sim$300~\pc, and having a thickness 
of $\sim$50~\pc. The amount of molecular mass in the CMZ, about 5(10$^7$)~\Msun~\citep{mor96},
would be consumed by star formation over relatively short timescales, 2-4(10$^8$)~yrs. Therefore, there must be a 
source of replenishment. The ring of molecular material which circumscribes
the CMZ at a galactocentric distance of 150 - 180~\pc\ is hypothesized to be material condensed 
into molecular form by shocks occurring along the innermost, non-self-intersecting X1 orbit in the
Galaxy's barred potential \citep{bin91,mor96}. This feature is fed from the outside by
gas migrating into the GC from the rest of the Galaxy. After shocking and condensing,
the gas continues its inward migration and moves onto x2 orbits inside the ring, where most
of the molecular material in the CMZ resides \citep{reg03}. 
The inflow rate of material from the ring can be estimated by dividing the mass in
the ring by the orbital period, 8(10$^6$)~\Msun/2(10$^7$)~yrs=0.4~\Msunyr. This is
an order of magnitude greater than the star formation rate estimated in this paper;
however, the mass budget also includes a term for mass lost through a thermally
driven wind (0.03 - 0.1~\Msunyr). Clearly, all of the terms in the mass budget have
errors that are large enough to permit the level of star formation claimed in
this paper.

\subsection{Relationship to Extragalactic Nuclear Populations}

Other than for the Milky Way, nuclei which harbor 
black holes and show no evidence of an AGN spectrum have
stellar populations consistent with formation in an ancient
burst \citep{mag98}. One must emphasize that 
black holes are more difficult to detect in galaxies with 
active star formation, so the kinematic sample is biased.
Within the Local Group, the nuclei of
M31 and M32 can be so characterized; while M32 may contain
some fraction of few \Gyr\ old stars, there is no evidence of recent
star formation in the M32 nucleus.
The unusual star formation history in the Galactic Center
may be related to bar-induced feeding of gas into the central
region, and subsequent star formation \citep{reg03}.

Our demonstration that the star formation history of the nuclear
population is continuous has additional implications, and raises several questions.
Has the black hole grown in mass
along with the stellar population, or was most of the mass of
the black hole in place, perhaps within a \Gyr\ of the Galaxy's
formation?  Further, how can a galaxy with a bulge population
well demonstrated to be as old as the oldest globular clusters
\citep{ort95,kk02,zoc03}, have formed at early times without simultaneously 
forming the bulk of the stars presently seen in the nuclear population?    

\acknowledgements
We thank Leo Girardi for graciously providing the Padova model isochrone files.
We acknowledge very useful discussions with Mike Regan, Paco Najarro, Bob Blum, Laurant Sjouwerman, 
and Jay Frogel. We also thank the Gemini North Science Verification Team for obtaining
the Galactic Center data used in this paper. Support for this work was provided by NASA 
through grant number GO-07364.01-96A and AR-08751.02-A from the Space Telescope Science 
Institute, which is operated by AURA, Inc., under NASA contract NAS5-26555. Mark Morris
is supported by NSF through AST9988397.

\clearpage
\begin{deluxetable}{lrrrrrrl}
\tabletypesize{\scriptsize}
\tablewidth{0pt}
\tablecaption{Log of Observations}
\tablehead{
\colhead{Field} &
\colhead{RA (1950)} &
\colhead{DEC (1950)} &
\colhead{Long.} &
\colhead{$\Delta$Long.\tablenotemark{a}} &
\colhead{Lat.} &
\colhead{$\Delta$Lat.\tablenotemark{a}} &
\colhead{Instrument} \\
\colhead{} &
\colhead{h m s} &
\colhead{\ \ \ \ \arcdeg \ \ \  \arcmin \ \ \ \arcsec} &
\colhead{degrees} &
\colhead{pc} &
\colhead{degrees} &
\colhead{pc} &
\colhead{}
}
\startdata
Quintuplet &	17 43 04.80	&$-$28 48 26.0	&0\fdg1664	&31.6	&$-$0\fdg0611	&$-$1.5 & HST/NICMOS \\
Arches	&17 42 39.90	&$-$28 48 13.0	&0\fdg1218	&25.4	&0\fdg0182	&9.5 & HST/NICMOS  \\
za	&17 42 34.47	&$-$28 45 57.1	&0\fdg1435	&28.4	&0\fdg0549	&14.6 & HST/NICMOS  \\ 
zb	&17 42 10.96	&$-$28 42 45.9	&0\fdg1435	&28.4	&0\fdg1559	&28.7 & HST/NICMOS  \\
zc	&17 41 47.47	&$-$28 39 34.4	&0\fdg1435	&28.4	&0\fdg2569	&42.9 & HST/NICMOS  \\ 
zd	&17 42 43.68	&$-$28 54 13.2	&0\fdg0439	&14.5	&$-$0\fdg0461	&0.5 & HST/NICMOS  \\
GC  & 17	42 30.00	&$-$28 53 00.0		&0\fdg0350	&12.7	&0\fdg0070	&7.4 &Lick/Gemini	\\
Gemini \#1 	& 17 42 29.22 &$-$28 59 19.4&$-$0\fdg0548	&0.13   &$-$0\fdg0455&0.11	&Gemini/AO/Hokupa'a+Quirc 	\\
Gemini \#2 	& 17 42 29.37	&$-$28 58 59.4&$-$0\fdg0498	&0.83   &$-$0\fdg0430&0.46	&Gemini/AO/Hokupa'a+Quirc 	\\
Gemini \#5 	& 17 42 30.73	&$-$28 59 19.5&$-$0\fdg0519&0.53 &$-$0\fdg0502	&$-$0.54	&Gemini/AO/Hokupa'a+Quirc 	\\
Gemini \#6 	& 17 42 30.82	&$-$28 58 59.5&$-$0\fdg0470&1.22 &$-$0\fdg0476	&$-$0.17	&Gemini/AO/Hokupa'a+Quirc 	\\
\enddata
\tablenotetext{a}{Values are with respect to Galactic Center, located at ($-$0\fdg0557,	$-$0\fdg0463).}
\end{deluxetable}

\clearpage

\begin{figure}
\epsscale{1.3}
\hspace{1.75in}
\hspace*{1.5in} 
\vskip .2in
\caption{Fields covered by observations. Labels are overplotted for the NICMOS
``z-fields'' (za, zb, zc, and zd), the cluster control fields (ac and qc), and the
Gemini/AO fields (the four fields nearest the GC label). The Lick field is marked by the large irregularly-shaped
box \citep{fig95a}. Coordinates for the fields are given in Table~1.
All figures may be obtained in the version of this paper at: http://www.stsci.edu/$\sim$figer/private/papers/gcsfrate/ms.ps.
\label{fig-fields}}
\end{figure}

\clearpage


\begin{figure}
\epsscale{1}
\hspace{3.75in}
\hspace*{4.5in} 
\vskip .2in
\caption{Observed color-magnitude diagrams for NICMOS data ({\it upper panel}), 
Gemini data ({\it middle panel}), and Lick data ({\it lower panel}). 
The ``red clump'' can be seen best in the NICMOS data, starting at H$-$K$\approx$1.5
and \mK$\approx$15, and continuing to the lower right; the extension toward the
lower right is a result of differential reddening dispersing the stars along the reddening vector.
All figures may be obtained in the version of this paper at: http://www.stsci.edu/$\sim$figer/private/papers/gcsfrate/ms.ps.
\label{fig-panels_cmd}}
\end{figure}

\clearpage

\begin{figure}
\epsscale{1}
\hspace{3.75in}
\hspace*{4.5in} 
\vskip .2in
\caption{Observed color-magnitude diagrams for individual NICMOS fields. Indications
of a red clump can be seen in all fields at locations that are dependent on the average
extinction for each field. For example, the red clump has a color of \mnh$-$\mnk$\approx$1.5 for
the zd field, the field that is nearest to the GC and that suffers the highest extinction of the six
fields. The zc field is furthest from the GC, and therefore has the lowest extinction and
a red clump that is relatively blue compared to the other fields. Note that the
gap between the main sequence (\mnk$\approx$19) and the red clump (\mnk$\approx$15.5)
is more populated for fields nearer to the GC, i.e.\ in the zd field, suggesting
a trend of younger stars toward the center.
All figures may be obtained in the version of this paper at: http://www.stsci.edu/$\sim$figer/private/papers/gcsfrate/ms.ps.
\label{fig-individualcmds}}
\end{figure}

\clearpage

\begin{figure}
\epsscale{1}
\hspace{3.75in}
\hspace*{4.5in} 
\vskip .2in
\caption{Dereddened luminosity functions for individual NICMOS fields, as extracted from
the color-magnitude diagrams in Figure~\ref{fig-individualcmds}; the data have not been
corrected for incompleteness. The red clump can
generally be seen near \mnk$\approx$12.5, although it is shifted toward brighter
magnitudes for fields closer to the GC. Evidence of relatively young stars 
(hundreds of Myr old) in the main sequence gap (e.g.\ Figure~\ref{fig-individualcmds}) 
can be seen in the luminosity function for the zd field. 
All figures may be obtained in the version of this paper at: http://www.stsci.edu/$\sim$figer/private/papers/gcsfrate/ms.ps.
\label{fig-individuallfs}}
\end{figure}

\clearpage

\begin{figure}
\epsscale{1.1}
\hspace{3.75in}
\hspace*{4.5in} 
\vskip .2in
\caption{Dereddened luminosity functions for the Galactic Center NICMOS ({\it dot-dashed}), 
Lick ({\it light}) fields, and Gemini/AO fields; the data have not been corrected
for incompleteness. The three luminosity functions match quite well over the magnitude
ranges for which they are complete. The Lick data are already seriously incomplete at \mK$_0\approx$9,
and the Gemini data begin to be incomplete at \mK$_0\approx$12, or roughly the magnitude
of the red clump. 
All figures may be obtained in the version of this paper at: http://www.stsci.edu/$\sim$figer/private/papers/gcsfrate/ms.ps.
\label{fig-total_lf0.0.za.zb.zc.zd.qca.qcb.qcc.qcd.aca.acb.acc.the.gem.black}}
\end{figure}

\clearpage

\begin{figure}
\epsscale{1}
\hspace{3.75in}
\hspace*{4.5in} 
\vskip .2in
\caption{Theoretical ({\it heavy}) and observed ({\it light}) luminosity function for the Galactic 
Center Lick field. The figure legends give the ages of the single starbursts assumed
for the models. We assume solar metalicity and the ``canonical'' mass-loss rates
in the Geneva models. The starburst models used in making this figure are used in
creating the summed luminosity functions for more complex star formation histories in
later figures. Note that we have not normalized the counts in this plot.
The plots show that the observed counts at the brightest magnitudes cannot be reproduced by
any single starburst. Note, also, that the counts at the bright end require the
presence of some young stars. 
All figures may be obtained in the version of this paper at: http://www.stsci.edu/$\sim$figer/private/papers/gcsfrate/ms.ps.
\label{fig-lfpanels0}}
\end{figure}

\clearpage

\begin{figure}
\epsscale{1}
\hspace{3.75in}
\hspace*{4.5in} 
\vskip .2in
\caption{Dereddened \mK-band magnitude of red clump for a model starburst population in the Galactic Center as a 
function of age, adapted from \citet{gir00}, assuming solar metallicity. 
All figures may be obtained in the version of this paper at: http://www.stsci.edu/$\sim$figer/private/papers/gcsfrate/ms.ps.
\label{fig-redclump}}
\end{figure}

\clearpage

\begin{figure}
\epsscale{1.1}
\hspace{3.75in}
\vskip .2in
\caption{Theoretical ({\it solid}) and observed luminosity functions for Baades Window ({\it dashed})
and the Galactic Center fields ({\it dot-dashed}). The theoretical curve was produced using
the Geneva models, assuming solar metalicity and a constant 
star formation rate of 0.07~\Msunyr\ from 10~\Gyr\ ago to the present. 
The Baade's window stars were arbitrarily scaled such that the counts roughly match the counts in the 
NICMOS data. Note the fainter red clump in the Baade's Window data.
All figures may be obtained in the version of this paper at: http://www.stsci.edu/$\sim$figer/private/papers/gcsfrate/ms.ps.
\label{fig-total_lf1.0c.za.zb.zc.zd.qca.qcb.qcc.qcd.aca.acb.acc.bw.black.cc}}
\end{figure}


\clearpage

\begin{figure}
\epsscale{1.1}
\hspace{3.75in}
\hspace*{4.5in} 
\vskip .2in
\caption{Model luminosity functions ({\it heavy, right}) as a function star formation scenario ({\it left}) for Z=Z$_\odot$ and the
canonical mass-loss rates of the Geneva models, compared to the observed luminosity function
for the Galactic Center fields ({\it light, right}). The model counts have been multiplied by the completeness
fraction of the observations. The models have been constrained to produce 2(10$^8$)~\Msun\ over
a circular area having r$<$30~\pc. Continuous star formation histories fit better than ancient bursts.
All figures may be obtained in the version of this paper at: http://www.stsci.edu/$\sim$figer/private/papers/gcsfrate/ms.ps.
\label{fig-sfpanels1.0c.za.zb.zc.zd.qca.qcb.qcc.qcd.aca.acb.acc.cc}}
\end{figure}

\begin{figure}
\epsscale{1.1}
\hspace{3.75in}
\hspace*{4.5in} 
\vskip .2in
\caption{Same as Figure~\ref{fig-sfpanels1.0c.za.zb.zc.zd.qca.qcb.qcc.qcd.aca.acb.acc.cc}, except using the
Padova models. 
All figures may be obtained in the version of this paper at: http://www.stsci.edu/$\sim$figer/private/papers/gcsfrate/ms.ps.
\label{fig-sfpanels1.0x.za.zb.zc.zd.qca.qcb.qcc.qcd.aca.acb.acc.cc}}
\end{figure}

\clearpage

\begin{figure}
\epsscale{1.1}
\hspace{3.75in}
\hspace*{4.5in} 
\vskip .2in
\caption{Model luminosity functions ({\it heavy}) as a function star formation scenario for Z=Z$_\odot$ and the canonical mass-loss rates of the Geneva models, compared to the observed luminosity function
for the Baade's Window fields ({\it light}). The models have not been scaled for mass, but rather
have been arbitrarily scaled along the
vertical axis to match the observed counts in the \mK=11.0 bin. 
All figures may be obtained in the version of this paper at: http://www.stsci.edu/$\sim$figer/private/papers/gcsfrate/ms.ps.
\label{fig-sfpanels1.0c.bw.auto}}
\end{figure}

\begin{figure}
\epsscale{1.1}
\hspace{3.75in}
\hspace*{4.5in} 
\vskip .2in
\caption{Same as Figure~\ref{fig-sfpanels1.0c.bw.auto}, except using the
Padova models. 
All figures may be obtained in the version of this paper at: http://www.stsci.edu/$\sim$figer/private/papers/gcsfrate/ms.ps.
\label{fig-sfpanels1.0x.bw.auto}}
\end{figure}

\clearpage

\begin{figure}
\epsscale{1.1}
\hspace{3.75in}
\hspace*{4.5in} 
\vskip .2in
\caption{Luminosity functions ({\it heavy, right}) as a function star formation scenario ({\it left}) for Z=2Z$_\odot$ and the
canonical mass-loss rates of the Geneva models, compared to the observed luminosity function
for the NGC6528 field ({\it light, right}). The models have not been scaled for mass, but rather
have been arbitrarily scaled along the
vertical axis to match the observed counts in the \mK=11.0 bin. 
All figures may be obtained in the version of this paper at: http://www.stsci.edu/$\sim$figer/private/papers/gcsfrate/ms.ps.
\label{fig-sfpanels2.0c.6528.auto}}
\end{figure}

\clearpage

\begin{figure}
\epsscale{1.1}
\hspace{3.75in}
\hspace*{4.5in} 
\vskip .2in
\caption{Same as Figure~\ref{fig-sfpanels2.0c.6528.auto}, except using the
Padova models. 
All figures may be obtained in the version of this paper at: http://www.stsci.edu/$\sim$figer/private/papers/gcsfrate/ms.ps.
\label{fig-sfpanels2.0x.6528.auto}}
\end{figure}

\clearpage

\begin{figure}
\epsscale{1.1}
\hspace{3.75in}
\hspace*{4.5in} 
\vskip .2in
\caption{Model luminosity functions ({\it heavy, right}) as a function star formation scenario ({\it left}) for Z=Z$_\odot$ and the
canonical mass-loss rates of the Geneva models, compared to the observed luminosity function
for the Galactic Center za field ({\it light, right}). The model counts have been multiplied by the completeness
fraction of the observations. The models have been constrained to produce 2(10$^8$)~\Msun\ over
a circular area having r$<$30~\pc. Continuous star formation histories fit better than ancient bursts. 
All figures may be obtained in the version of this paper at: http://www.stsci.edu/$\sim$figer/private/papers/gcsfrate/ms.ps.
\label{fig-sfpanels1.0c.za.cc}}
\end{figure}

\clearpage

\begin{figure}
\epsscale{1.1}
\hspace{3.75in}
\hspace*{4.5in} 
\vskip .2in
\caption{Same as Figure~\ref{fig-sfpanels1.0c.za.cc}, except using the
Padova models. 
All figures may be obtained in the version of this paper at: http://www.stsci.edu/$\sim$figer/private/papers/gcsfrate/ms.ps.
\label{fig-sfpanels1.0x.za.cc}}
\end{figure}

\clearpage

\begin{figure}
\epsscale{1.1}
\hspace{3.75in}
\hspace*{4.5in} 
\vskip .2in
\caption{Model luminosity functions ({\it heavy, right}) as a function star formation scenario ({\it left}) for Z=Z$_\odot$ and the
canonical mass-loss rates of the Geneva models, compared to the observed luminosity function
for the Galactic Center zb field ({\it light, right}). The model counts have been multiplied by the completeness
fraction of the observations. The models have been constrained to produce 2(10$^8$)~\Msun\ over
a circular area having r$<$30~\pc. Continuous star formation histories fit better than ancient bursts. 
All figures may be obtained in the version of this paper at: http://www.stsci.edu/$\sim$figer/private/papers/gcsfrate/ms.ps.
\label{fig-sfpanels1.0c.zb.cc}}
\end{figure}

\clearpage

\begin{figure}
\epsscale{1.1}
\hspace{3.75in}
\hspace*{4.5in} 
\vskip .2in
\caption{Same as Figure~\ref{fig-sfpanels1.0c.zb.cc}, except using the
Padova models. 
All figures may be obtained in the version of this paper at: http://www.stsci.edu/$\sim$figer/private/papers/gcsfrate/ms.ps.
\label{fig-sfpanels1.0x.zb.cc}}
\end{figure}

\clearpage

\begin{figure}
\epsscale{1.1}
\hspace{3.75in}
\hspace*{4.5in} 
\vskip .2in
\caption{Model luminosity functions ({\it heavy, right}) as a function star formation scenario ({\it left}) for Z=Z$_\odot$ and the
canonical mass-loss rates of the Geneva models, compared to the observed luminosity function
for the Galactic Center zc field ({\it light, right}). The model counts have been multiplied by the completeness
fraction of the observations. The models have been constrained to produce 2(10$^8$)~\Msun\ over
a circular area having r$<$30~\pc. Continuous star formation histories fit better than ancient bursts. 
All figures may be obtained in the version of this paper at: http://www.stsci.edu/$\sim$figer/private/papers/gcsfrate/ms.ps.
\label{fig-sfpanels1.0c.zc.cc}}
\end{figure}

\clearpage

\begin{figure}
\epsscale{1.1}
\hspace{3.75in}
\hspace*{4.5in} 
\vskip .2in
\caption{Same as Figure~\ref{fig-sfpanels1.0c.zc.cc}, except using the
Padova models. 
All figures may be obtained in the version of this paper at: http://www.stsci.edu/$\sim$figer/private/papers/gcsfrate/ms.ps.
\label{fig-sfpanels1.0x.zc.cc}}
\end{figure}

\clearpage

\begin{figure}
\epsscale{1.1}
\hspace{3.75in}
\hspace*{4.5in} 
\vskip .2in
\caption{Model luminosity functions ({\it heavy, right}) as a function star formation scenario ({\it left}) for Z=Z$_\odot$ and the
canonical mass-loss rates of the Geneva models, compared to the observed luminosity function
for the Galactic Center zd field ({\it light, right}). The model counts have been multiplied by the completeness
fraction of the observations. The models have been constrained to produce 2(10$^8$)~\Msun\ over
a circular area having r$<$30~\pc. Continuous star formation histories fit better than ancient bursts. 
All figures may be obtained in the version of this paper at: http://www.stsci.edu/$\sim$figer/private/papers/gcsfrate/ms.ps.
\label{fig-sfpanels1.0c.zd.cc}}
\end{figure}

\clearpage

\begin{figure}
\epsscale{1.1}
\hspace{3.75in}
\hspace*{4.5in} 
\vskip .2in
\caption{Same as Figure~\ref{fig-sfpanels1.0c.zd.cc}, except using the Padova models. 
All figures may be obtained in the version of this paper at: http://www.stsci.edu/$\sim$figer/private/papers/gcsfrate/ms.ps.
\label{fig-sfpanels1.0x.zd.cc}}
\end{figure}

\clearpage

\begin{figure}
\epsscale{1.1}
\hspace{3.75in}
\hspace*{4.5in} 
\vskip .2in
\caption{Extinction map (\AK), as inferred from the Lick data \citep{fig95a}. Sgr~A* is located at (0,0).
The offsets are in arcseconds along RA and Dec. See text for discussion.
All figures may be obtained in the version of this paper at: http://www.stsci.edu/$\sim$figer/private/papers/gcsfrate/ms.ps.
\label{fig-extinction}}
\end{figure}

\end{document}